\documentstyle[11pt,aaspp4,epsfig]{article}
\slugcomment{To appear in Astrophysical Journal Letters}

\begin{document}

\title{Soft Phase Lags of Pulsed Emission from the Millisecond X-ray Pulsar SAX J1808.4-3658}
\author{Wei Cui\altaffilmark{1}, Edward H.~Morgan\altaffilmark{1}, \& Lev G.~Titarchuk\altaffilmark{2,3}}
\altaffiltext{1}{Center for Space Research, Massachusetts Institute of Technology, Cambridge, MA 02139; cui@space.mit.edu; ehm@space.mit.edu.}
\altaffiltext{2}{NASA/Goddard Space Flight Center, Code 661, Greenbelt, MD 20771; titarchuk@lheavx.gsfc.nasa.gov.}
\altaffiltext{3}{Also George Mason University, Institute for Computational Sciences and Informatics.}

\begin{abstract}
We report the discovery of phase shifts between X-ray pulses at different 
energies in the newly discovered millisecond (ms) X-ray pulsar SAX 
J1808.4-3658. The results show that low-energy pulses lag high-energy pulses 
by as much as $\sim$0.2 ms (or $\sim$8\% of the pulse period). The measurements
were made in two different ways: (1) computing cross power spectra between 
different energy bands, and (2) cross-correlating the folded pulse profiles 
in different energy bands; consistent results were obtained. We speculate that
the observed {\it soft lags} might be related to the lateral expansion and 
subsequent cooling of a ``hot spot'' on the neutron star surface in which the 
pulsed X-ray emission originates. Also presented is the possibility of 
producing soft lags via Compton {\it down} scattering of hard X-ray photons 
from the hot spot in the cool surrounding atmosphere. We will discuss 
possible X-ray production mechanisms for SAX J1808.4-3658 and constraints on 
the emission environment, based on the observed soft lags, pulse profiles, 
and energy spectrum. 
\end{abstract}


\section{Introduction}
Low-mass X-ray binaries (LMXBs) that contain a weakly magnetized neutron star 
are thought to be the progenitors of millisecond (ms) radio pulsars (see
review by Bhattacharya \& van den Heuvel 1991). Over the past two 
decades, extensive searches have been made for signatures of a rapidly 
spinning neutron star in such LMXBs (Leahy et al. 1983; Mereghetti \& Grindlay
1987; Wood et al. 1991; Vaughan et al. 1994). No coherent ms X-ray pulsation 
has been detected in these attempts. Near coherent ms oscillations have been 
observed in several sources, but {\it only} during thermonuclear (type I) 
X-ray bursts (Strohmayer et al. 1996; Zhang et al. 1996; Smith et al. 1997; 
Strohmayer et al. 1997; Zhang et al. 1998). They are interpreted as X-ray
intensity being modulated at the spin period of the neutron star. Such an 
interpretation is, however, still model dependent. The lack of coherent 
pulsation in the persistent emission of LMXBs can perhaps be attributed to 
the low magnetic field believed to exist in these systems, compared to that in
typical X-ray pulsars, or the smearing of pulsar signal by such effects as 
gravitational lensing (Wood et al. 1988; Meszaros et al. 1988) or scattering 
(Brainerd \& Lamb 1987; Kylafis \& Klimmis 1987; Wang \& Schlickeiser 1987; 
Bussard et al. 1988).

Very recently, a coherently pulsed X-ray signal was revealed at a period of 
$\sim$2.49 ms in the observations of XTE J1808-369 with the {\it Proportional 
Counter Array} (PCA) aboard the {\it Rossi X-ray Timing Explorer} (RXTE) 
(Wijnands \& van der Klis 1998) during its recent outburst (Marshall 1998). 
Raster scans were made to locate this newly discovered source, and the results
imply that its position is consistent with that of SAX J1808.4-3658 (Marshall 
1998), which was discovered by BeppoSAX in the midst of a previous outburst 
(in 't Zand et al. 1998). Moreover, the results from subsequent timing 
analysis of 
the RXTE/PCA observations seem to favor the BeppoSAX coordinates (Chakrabarty 
\& Morgan 1998). During the previous outburst, BeppoSAX detected Type I X-ray 
bursts from SAX J1808.4-3658 (in 't Zand et al. 1998). The bursting activity 
generally indicates the presence of a weakly magnetized neutron star in a 
binary system (Lewin et al. 1995, and references therein). The binary nature 
of SAX J1808.4-3658 was firmly established with the detection of a 2-hour 
orbital period (Chakrabarty \& Morgan 1998), as well as with the optical 
identification of the companion star (Roche et al 1998). 

The observed X-ray spectrum of SAX J1808.4-3658 is unusually hard for an X-ray
pulsar (Gilfanov et al. 1998;
Heindl \& Smith 1998). It can be characterized by a Comptonized spectrum from 
a region of electron temperature $kT_e = 22$ keV and optical depth $\tau = 4$ 
(or $2$) for a spherical (or slab) geometry (Heindl \& Smith 1998; see 
Titarchuk 1994 for a discussion on different scattering geometries). The 
process of inverse Comptonization would cause high-energy photons lag 
low-energy photons (see, e.g., Sunyaev \& Titarchuk 1980, hereafter ST80). 
This consideration prompted us to search for any {\it hard} phase lags of 
X-ray emission from SAX J1808.4-3658. In this Letter, we report the discovery 
of significant {\it soft} phase lags of the pulsed emission, which are rather 
unexpected. In the 
framework of Comptonization models, the soft lags can be readily explained 
by Compton {\it down} scattering of high-energy photons, which would indicate 
the importance of the re-processing of hard radiation from the neutron star 
surface by 
a cool surrounding atmosphere. We will present arguments for and against this 
interpretation, and will also suggest an alternative scenario. 

\section{Data Analysis and Results}
The data used for this study come from 19 PCA observations (out of a total of 
21; the longer of the two observations was selected for April 18 and May 2). 
In particular, the {\it Event} mode data with $\sim 122\mu s$ timing 
resolution and 64 energy bands were selected (except for the April 13 
observation in which the {\it goodXenon} 
modes were used) to facilitate high-resolution timing analysis with a moderate
energy resolution. A mixture of short and long pointed observations were 
conducted, with the effective exposure time ranging from $\sim$1.4 ks to 
$\sim$25 ks. 

We carried out spectral analysis, using the {\it Standard 2} data (with 
16-second timing resolution). Limited by the calibration uncertainties in
the PCA response matrices, we selected only 75 out of 129 energy channels
to cover an energy range 2.5--30 keV. Throughout the entire period, the 
observed X-ray spectrum maintains a rough power-law shape of photon index 
$\sim$2 (see also Gilfanov et al. 1998 and Heindl \& Smith 1998). The 
addition of a soft component (e.g., 
blackbody) improves the model fit significantly in terms of $\chi^2$ 
statistics, confirming the reported soft excess (Heindl \& Smith 1998). 
For each observation, we computed the observed X-ray flux by taking into 
account of the PCA pointing offset (1.4\arcmin\ except for 
the first observation where the offset is $\sim$12.3\arcmin). Fig.~1 shows the
decaying of the outburst, during which the X-ray flux varied by more than 
two orders of magnitude. Following the initial phase of an exponential decay,
the flux started to drop precipitously around April 26. At the lowest fluxes, 
source confusion and background subtraction become serious problems for 
analyzing PCA observations. A raster scan was purposefully planned and carried 
out at the beginning of the last observation (on May 6), which showed
that the detected X-ray emission was indeed from SAX J1808.4-3658 and no 
apparent contaminating sources were present in the PCA field-of-view. 

Because of large Doppler effects due to the orbital motion for SAX 
J1808.4-3658, it is more convenient to adopt a reference frame centering on
the neutron star and rotating with the binary motion. After 
correcting photon arrival times for RXTE's motion with respect to the 
barycenter of the solar system,
we proceeded to take out the effects of binary motion by using the 
measured binary parameters (Chakrabarty \& Morgan 1998). 

We folded the corrected light curves (with background subtracted) at 
the pulse period in several energy bands. The measured fractional RMS pulse 
fraction in the summed band (2--30 keV) is also plotted in Fig.~1. An 
anti-correlation is apparent in the figure
between the X-ray flux and fractional pulse amplitude. When the flux dropped 
below $2 \times 10^{-11}\mbox{ }ergs\mbox{ }cm^{-2}\mbox{ }s^{-1}$ on May 2, 
the pulse signal became barely detectable. The detection significance 
jumped up the next day, as the source flux increased by roughly a factor of 3,
and dropped again in the last observation, as the source flux decreased again.

Typical pulse profiles are shown in Fig.~2 (also see Wijnands \& van der Klis
1998), taking the results from the April 
23 observation (with an exposure time $\sim$17 ks). The pulse profiles can be 
modeled adequately 
by a single sine function, but the fits are much improved with the inclusion
of contributions from high-order harmonics. To improve statistics, we 
combined data from the observations between April 11-29 (as indicated in 
Fig.~1, with a total exposure time $\sim$150 ks), when 
the pulse signal is detected with high significance, to obtain the average 
pulse profiles. By modeling these profiles, we measured average fractional 
RMS pulse fractions for the fundamental component and high-order
harmonics. The results are summarized in Fig.~3. For the April 11 observation,
our results are in agreement with those derived by Wijnands \& van der Klis 
(1998). The fractional pulse 
amplitude shows an initially decreasing trend with energy for the fundamental 
component, but the opposite for the first harmonic (the error bars are quite
large for the second harmonic). It seems to level off for both components 
above $\sim$10 keV.
 
Finally, for each 128-second data segment we constructed a power-density 
spectrum (PDS) and a cross-power spectrum (CPS) between the 2-3 keV band 
and each of several higher energy bands. Except for the observations on May 
2 and 6 (in Fig.~1), the pulse signal shows up prominently as a peak in the 
PDS. There is also significant broad-band power extending up to a few tenth 
Hz before dropping off toward higher frequencies. Individual 
CPSs are then properly weighted and co-added to obtain the average CPS for 
the observations between April 11-29. The average phase difference between 
X-ray pulses in two energy bands is directly derived from the average 
CPS. The results imply that low-energy pulses {\it lag} high-energy pulses, 
as shown in Fig.~4. Actually, the soft lags are also apparent in Fig.~2. By 
cross-correlating the folded pulse profiles at different energies, we also 
derived soft phase lags, which are consistent with those derived from the
average CPS.

\section{Discussion}
For SAX J1808.4-3658, the observed characteristics of the pulsed X-ray emission
(pulse profiles and phase lags) and the overall energy spectrum can provide 
useful insights into X-ray production processes and the emission environment. 
The pulsed X-ray emission was detected in all 
observations. Integrating the best-fit Comptonization model (ST80) for the 
May 6 observation when the minimum flux was reached, we derived
a bolometric luminosty $\sim 1.0\times 10^{35}\mbox{ }ergs\mbox{ }s^{-1}$ 
(assuming a source distance 4 kpc; in 't Zand et al. 1998). The lack of 
centrifugal inhibition of accretion flows to the magnetic poles at such a low 
corresponding accreton rate implies the presence of a very weak magnetic field 
in the system ($\lesssim 0.4-1.3\times 10^8$ G; cf. Wijnands \& van der Klis 
1998). Such a weak field is quite 
unusual for an accreting X-ray pulsar but is certainly consistent with our 
current knowledge about type~I X-ray bursters. 

If the soft phase lag is due to Compton {\it down} scattering of pulsed hard 
X-ray emission by relatively cool medium, the observed pulse profile would be 
more sinusoidal at low energies, which is indeed observed (see, e.g., Fig.~3).
The leveling-off of the 
fractional pulse amplitudes seems to indicate the fact that the intrinsic 
values are approached at high energies. This then implies that the intrinsic 
pulse profile is highly sinusoidal.

The size of the scattering medium can be constrained by the observed soft 
lags. For simplicity, we assume that the input photons are monochromatic with
energy $E_i$. The 
photons that emerge from the cloud with {\it lower} energy $E_l$ arrive at a 
distant observer later than those with {\it higher} energy $E_h$. The delay 
in the arrival time, $\delta t$, is given by $\sim \Delta u l/c$, where 
$\Delta u$ is the difference in the average number of scatters experienced by 
seed photons before emerging with energies $E_l$ and $E_h$; and $l$ is 
the photon mean free path. The electron temperature of the cloud is likely to 
be a fraction of keV, as required by the energetics in the vicinity of the
neutron star. For cases where $kT_e \ll E_i$, the average fractional energy 
loss of input photons after each scatter is nearly independent of $T_e$ and 
is given by $\Delta E/E \approx -E/m_e c^2$, where $m_e$ is the electron mass.
Integrating over multiple scatters, we have 
$\Delta u = m_e c^2 (1/E_l - 1/E_h)$. Substituting this result into the 
expression for $\delta t$ gives
\begin{equation}
\delta t \sim \frac{r}{c\tau} (\frac{m_e c^2}{E_l} - \frac{m_e c^2}{E_h}),
\end{equation}
where $r$ is the radius of a spherical ``cloud'' into the center of which 
input photons are injected; and $\tau = r/l$ is its optical depth. The 
measured 
soft lag scales very roughly as $E^{-1}$, as shown in Fig.~4, and it seems 
to level off above 10 keV. In reality, however, the situation is much more 
complicated. The input photons may be distributed over a large energy range 
as well as over an extended region spatially. Moreover, the analysis only 
deals with broad energy bands rather than the energy of individual photons. 
Convolved with input photon distribution both in energy and space, the 
soft-lag 
plateau seems to imply that the ``effective'' energy of input photons is 
$\sim$10 
keV. Because it takes more than 100 scatterings for a 10-keV photon to reach 
the reference band and the average number of scatterings is on the order of
$\tau^2$ (ST80), the cloud must be quite large ($\tau \gtrsim 10$). It 
would then be imperative that the hot spot is viewed directly, since the
scattering process would significantly soften the spectrum (ST80). Hard 
photons from the hot spot are down-scattered in the cool surroundings to 
produce the observed soft lags and, perhaps, also the soft excess observed. A 
fit to the initial portion of the curve with equation (1) yields 
$r/c\tau \simeq 1.75 \mu s$, hence, $r \sim 0.5 \tau$ km. 
Therefore, the cloud is a few kilometers in size. Given the compactness of 
the binary system for SAX J1808.4-3658, significant X-ray heating of the 
companion star is expected (see discussion in Chakrabarty \& Morgan 1998). As 
a result, the mass loss from the companion star is much enhanced, perhaps 
forming a relatively dense wind that scatters hard X-rays originating in the 
vicinity of the neutron star. It is, however, not clear how to produce a 
``hole'' through such an extended cloud toward the hot spot. 

Alternatively, the soft lags might be caused by hydrodynamical propagation 
of the hot spot over the neutron star surface. Compared to typical X-ray 
pulsars, SAX 
J1808.4-3658 only contains a very weakly magnetized neutron star. The magnetic 
confinement of plasma in the hot spot is therefore relatively weak. 
Consequently, the plasma could spread out over the neutron star surface 
relatively 
easily, at hydrodynamical velocities of the order of sound speed. During
the lateral expansion, the outskirts of the hot spot cools down --- the 
temperature is approximately inversely proportional to the square root of 
the spot size. For a circular spot of radius 0.5 km and of temperature 25 
keV, the propagation time scale is roughly $3\times 10^{-4}$ s. This process 
could therefore account for the observed soft lags, if the soft photons which 
lag originate in the cool outskirts of the expanding hot spot. Assuming 
$E\propto kT$, since $kT\propto r^{-1/2}$ and $r \sim c_s \delta t$, we get
$\delta t\propto E^{-2}$, which is not inconsistent with the data (see the
initial portion of the curve shown in Fig.~4). In the context of this model, 
the pulse profile is also expected to be smoother at low energies because the 
softer photons come from a larger area (and thus are more integrated).

The X-ray pulsation is detected at high energies (see Fig.~2), implying 
that the hot spot produces very hard photons. For LMXBs that contain a weakly 
magnetized neutron star the physical processes for producing hard X-ray 
radiation have been discussed in literature since late sixties. Zeldovich and 
Shakura (1969) presented a model where the gravitational energy of matter 
accreted onto the neutron star is released in a thin layer above the neutron 
star surface. Variations of this idea have also been proposed and formulated 
quantitatively in detail models (see, e.g., Alme \& Wilson 1973 and Basko \& 
Sunyaev 1975a for accretion in the presence of magnetic field; Turolla et al. 
1994 for spherical accretion; and Klu\`{z}niak \& Wilson 1991 for ``gap 
accretion''). The deep layers of the neutron star atmosphere are heated
by the outer layer and produce soft thermal photons. The soft photons are 
subsequently Compton upscattered by hot electrons in the outer layer and form 
a thermal Comptonized spectrum (ST80). Titarchuk et al. (1998) verified 
Zeldovich \& Shakura's calculation and showed that for a layer of optical 
depth a few {\it the product of the optical depth and plasma temperature is 
almost invariant}. Therefore, the Comptonized spectrum from the layer 
maintains roughly the same shape as long as mass accretion rate is about 10\%
of the Eddington limit or less. Note that the {\it hard} lags of the 
emission due to Compton upscattering are negligible because of the compactness
of the region. For SAX J1808.4-3658, the mass accretion rate is less than 
$10^{-9}\mbox{ }M_{\odot}\mbox{ }yr^{-1}$ throughout the recent outburst, so 
the optical depth of the accreton column is always relatively small. The 
combination 
of low magnetic field and thin accretion column can easily result in hot spots
with a temperature $kT_e \sim 20$ keV, as observed (Heindl \& Smith 1998). 
Also observed is the expected constancy of the X-ray spectral shape during the
outburst (Gilfanov et al. 1998). It is worth noting that
similar models cannot be applied to typical X-ray pulsars where magnetic 
field is strong. In those cases, the proton energy lose due to Coulomb 
collisions becomes negligible compared to that due to nucleon-nucleon 
collisions, and thus the mean free path for energy release can become quite 
large (Basko \& Sunyaev 1975b). As a result, the temperature of the 
Comptonizing layer is relatively low, so the spectrum is usually soft. 

\acknowledgments
We thank Frank Marshall for useful comments on an early draft of the paper
and Rudy Wijnands, the referee, for helpful suggestions that significantly
improved the presentation. 
We also wish to acknowledge stimulating conversations with Wlodek Klu\'{z}niak,
Demos Kazanas, Shuang Nan Zhang, and Deepto Chakrabarty. This work is supported
in part by NASA through Contract NAS5-30612.

\clearpage
\begin{figure}
\psfig{figure=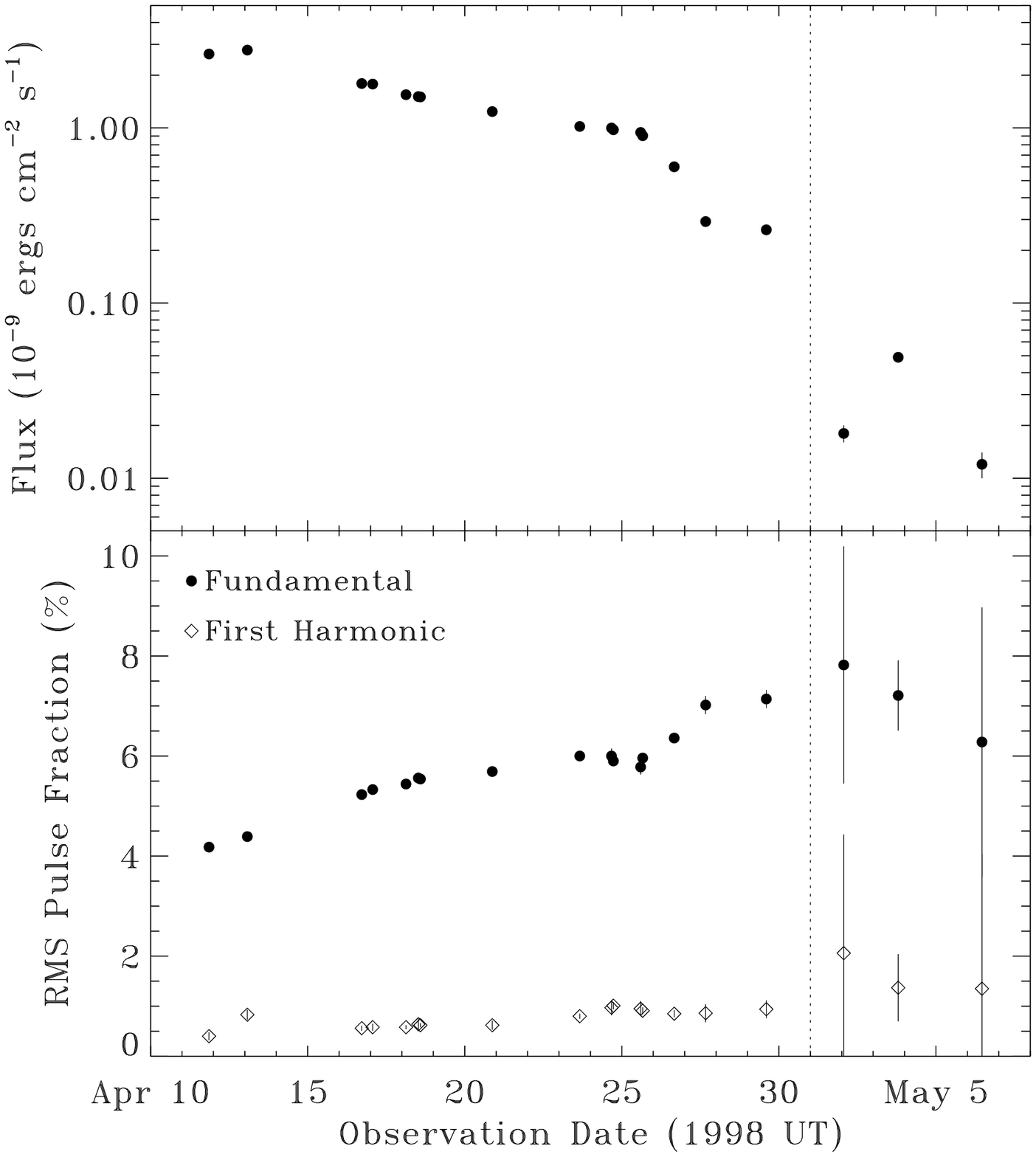,width=4in}
\caption{X-ray flux and fractional RMS pulse amplitude. The measurements were 
made in the 2--30 keV band. Note that the dotted line indicates roughly the
start of a period when the pulse signal was not detected with high 
significance. }
\end{figure}

\begin{figure}
\psfig{figure=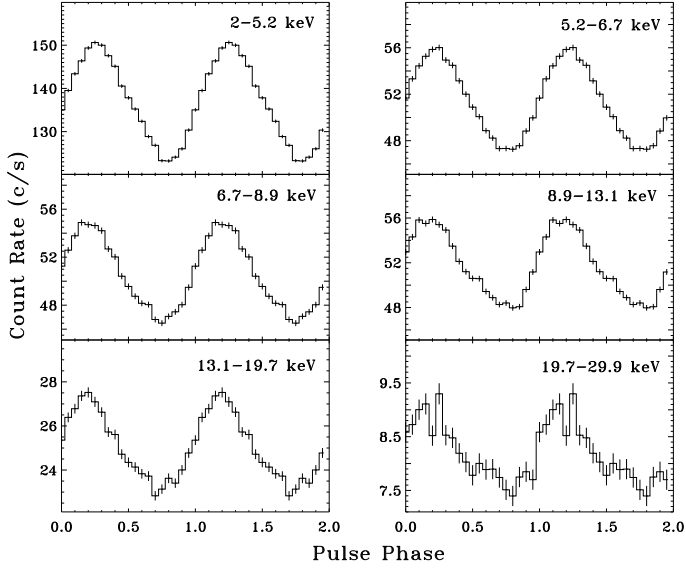,width=4in}
\caption{Sample pulse profiles. Folded are the light curves in six energy 
bands for the observation taken on 23 April 1998 (see Fig.~1). Note that 
each profile is repeated in two cycles for clarity. }
\end{figure}

\begin{figure}
\psfig{figure=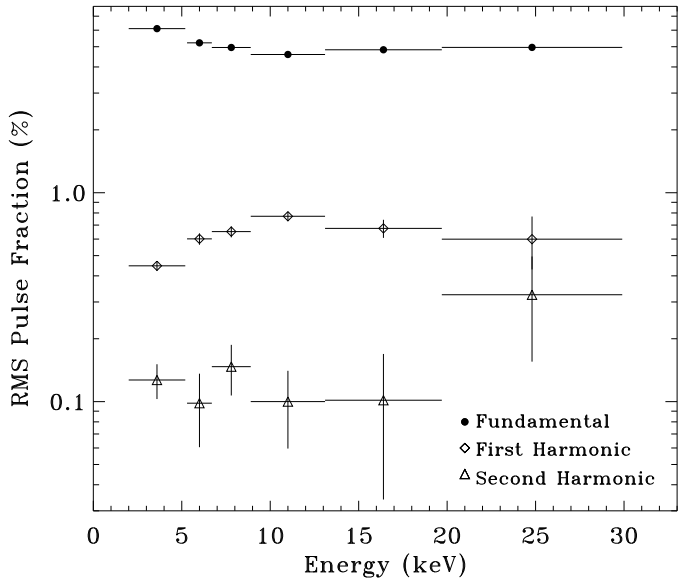,width=4in}
\caption{Energy dependence of measured pulse amplitudes. The results were
obtained by averaging over data from the observations between April 11-29, 
as indicated in 
Fig.~1. Note that the error bars for the fundamental component are totally 
negligible compared to the size of the symbols used and therefore the 
initial decreasing trend of the pulse amplitude is highly significant. }
\end{figure}

\begin{figure}
\psfig{figure=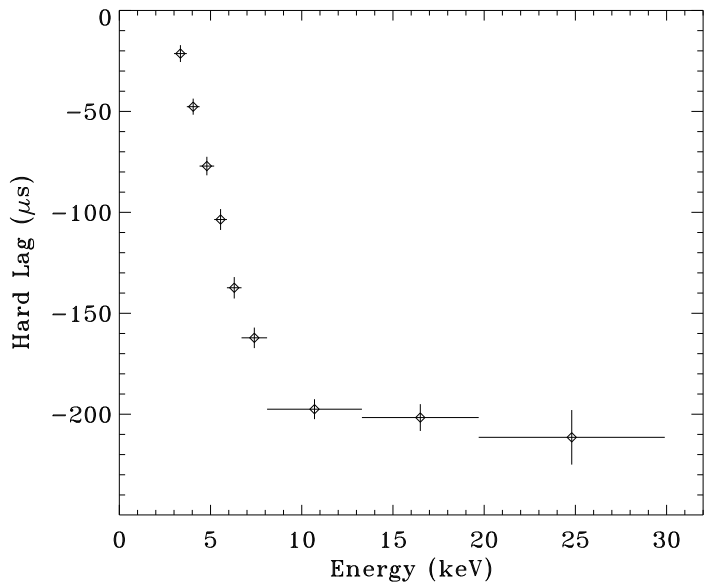,width=4in}
\caption{Measured hard X-ray lags with respect to the 2--3 keV band. Note 
that the data points are plotted arbitrarily in the middle of each energy 
band. The results have been averaged over the observations between April 
11-29. The negative values emphasize the fact that hard X-rays actually 
{\it lead} soft X-rays. }
\end{figure}

\end{document}